\documentclass[conference]{IEEEtran}
\IEEEoverridecommandlockouts

\usepackage{cite}
\usepackage{amsmath,amssymb,amsfonts}
\usepackage{algorithmic}
\usepackage{graphicx}
\usepackage{textcomp}
\usepackage{xcolor}

\newtheorem{definition}{Definition}
\def\BibTeX{{\rm B\kern-.05em{\sc i\kern-.025em b}\kern-.08em
    T\kern-.1667em\lower.7ex\hbox{E}\kern-.125emX}}
\newcommand{\transp}{\text{T}}

\ifCLASSOPTIONcompsoc \usepackage[caption=false,font=normalsize,labelfon
t=sf,textfont=sf]{subfig}
\else
\usepackage[caption=false,font=footnotesize]{subfi g}
\fi

\newcommand{\mathclap}[1]{\text{\hbox to 0pt{\hss$\mathsurround=0pt#1$\hss}}}

\begin{document}

\title{Sparse wavefield reconstruction and denoising with boostlets\\
\thanks{This work is financially supported by the Swedish Research Council, grant agreement No. 2020-04668.}
}

\author{\IEEEauthorblockN{Elias Zea}
\IEEEauthorblockA{\textit{Department of Engineering Mechanics} \\
\textit{KTH Royal Institute of Technology}\\
Stockholm, Sweden \\
zea@kth.se}
\and
\IEEEauthorblockN{Marco Laudato}
\IEEEauthorblockA{\textit{Department of Engineering Mechanics} \\
\textit{KTH Royal Institute of Technology}\\
Stockholm, Sweden \\
laudato@kth.se}
\and
\IEEEauthorblockN{Joakim Andén}
\IEEEauthorblockA{\textit{Department of Mathematics} \\
\textit{KTH Royal Institute of Technology}\\
Stockholm, Sweden \\
janden@kth.se}
}

\maketitle

\begin{abstract}
Boostlets are spatiotemporal functions that decompose nondispersive wavefields into a collection of localized waveforms parametrized by dilations, hyperbolic rotations, and translations.
We study the sparsity properties of boostlets and find that the resulting decompositions are significantly sparser than those of other state-of-the-art representation systems, such as wavelets and shearlets. 
This translates into improved denoising performance when hard-thresholding the boostlet coefficients. 
The results suggest that boostlets offer a natural framework for sparsely decomposing wavefields in unified space--time.
\end{abstract}

\begin{IEEEkeywords}
wavefields, sparse reconstruction, denoising, multi-scale representations, boostlets. 
\end{IEEEkeywords}

\section{Introduction}

In 1946, Dennis Gabor developed a theory for time-frequency analysis~\cite{Gabor1946}, envisioning that a family of ``atoms'' could detect transient events more effectively than Fourier transforms.
This work inspired the development of the wavelet transform by Grossmann, Morlet, Daubechies, and Meyer~\cite{Grossmann1984, Daubechies1992, Meyer1993}.
What is striking about wavelets is their natural, physical relationship to localized singularities in high-dimensional spaces.
Indeed, wavelets are fundamental mathematical objects that transcend applications in problems ranging from gravitational wave detection~\cite{klimenko2004wavelet} to JPEG2000 compression~\cite{unser2003mathematical}.

Building on this foundation, many other multi-scale directional representation systems have been developed to represent signal classes with characteristic features, such as point singularities, edges, wavefronts, etc.
These representation systems can be learned from data~\cite{Bianco2017, Hahmann2021, Zea2021} or can be handcrafted in accordance with certain characteristics of the signals in question, such as the case of isotropic~\cite{Daubechies1992} and directional wavelets~\cite{Antoine1999}, ridgelets~\cite{Candes1999}, curvelets~\cite{Candes1999b, Candes2004}, contourlets~\cite{Do2005}, shearlets~\cite{Guo2006}, and $\alpha$-molecules~\cite{Grohs2016}. 

In the field of acoustics, this development has been particularly beneficial in combination with compressed sensing techniques~\cite{Candes2006, Donoho2006}, which has resulted in a valuable reduction in the number of sensors (e.g., microphones) required for various data acquisition and signal processing tasks.
This has led to dictionaries specifically adapted to acoustic signals, such as virtual monopole sources~\cite{Mignot2013, Antonello2017} as well as modal~\cite{Haneda1999, Das2021}, and plane-wave expansions~\cite{Mignot2011, Mignot2014, Jin2015, Verburg2018}.



One important aspect of acoustic signal processing is to consider the recorded pressure signal as a function of time and space, referred to as a \emph{wavefield} recording.
In this direction, several researchers have employed curvelet frames~\cite{Herrmann2008} and cone-adapted shearlet dictionaries~\cite{Zea2019} to interpolate these wavefields in unified space--time.
Others have addressed sampling and interpolating acoustic wavefields in the spatial domain~\cite{Ajdler2006, Pinto2010, Pinto2011, Mignot2011, Berkhout1993, Berkhout1997, Berkhout1999}.
There are similar works on seismology~\cite{Yarman2020} and dynamic tomography~\cite{Bubba2023} applied to signals in space--time.

One important contribution in this line of research is that of the directional filter banks introduced by Pinto and Vetterli~\cite{Pinto2010, Pinto2011}, whereby Gabor's works on time--frequency analysis~\cite{Gabor1946} are extended to space--time--frequency analysis.
The authors coined the \emph{short space--time Fourier transform} to emphasize such an extension to space--time.
The underlying idea of such a transform is to apply a spatiotemporal window to signals recorded with a uniform microphone array.
These windows are then designed to capture the directionality (i.e., phase speed) of far- and near-field waves in space--time using a collection of directional filter banks. 

A central aspect of these filter banks is the bandlimitedness of phase velocities in acoustic wavefront data and the separation between far-field and near-field spectral components.
However, they do not properly account for the frequency-dependent behavior exhibited by the spectral attributes of the sources and boundary conditions in the measurement environment.
How to model such a frequency behavior together with the phase velocities has motivated the development of the continuous \emph{boostlet transform}~\cite{Zea2021, zeaBoostlets}, which leverages the application of groups used in relativity theory to form a natural dictionary for decomposing acoustic wavefields in space--time.

In this work, we study the sparsity properties of the resulting boostlet decomposition and compare it to other multi-scale decomposition methods, such as Daubechies45 wavelets and shearlets.
We see that, for acoustic wavefields, the boostlets yield a more compact representation, providing a higher-accuracy reconstruction from fewer coefficients. This results in improved denoising performance via hard thresholding of the boostlet coefficients. These results suggest that boostlets are natural representations of wavefields in unified space--time. 

\section{Boostlet preliminaries}

\subsection{Continuous boostlets}
Boostlets are wavelet-based functions parametrized with isotropic dilations, hyperbolic rotations, and translations in space-time. The powerful machinery behind this parametrization is the action of the Poincaré group with isotropic dilations, also known as the Weyl group in the theory of conformal fields~\cite{Weyl1993}. Let us make this parametrization mathematically precise with the following definitions.

\begin{definition}[Dilation and boost matrices]
Define a dilation matrix $D_a$ and a boost matrix $B_\theta$ acting on space--time vectors $\varsigma = (x,t)^\transp \in \mathbb{R}^2$ as
\begin{equation}
    D_a = \begin{pmatrix} a & 0 \\ 0 & a \end{pmatrix}, 
    \,\,\, B_\theta = \begin{pmatrix} \cosh \theta & -\sinh \theta \\ -\sinh \theta & \cosh \theta \end{pmatrix},
\end{equation}
with dilation parameter $a \in \mathbb{R}^+$, and Lorentz boost (hyperbolic rotation) parameter $\theta \in \mathbb{R}$.
It is often convenient to combine these two transformations into a single dilation--boost matrix $M_{a,\theta}$ given by
\begin{equation}
    M_{a,\theta} = D_a B_\theta = B_\theta D_a = \begin{pmatrix} a \cosh \theta & -a \sinh \theta \\ -a \sinh \theta & a \cosh \theta \end{pmatrix}.
\end{equation}
\end{definition}
While the dilation performs a simple scaling in both time and space, the boost operator performs a hyperbolic rotation in space--time.
In terms of acoustic waves, the latter can be seen as modifying its phase velocity.

Given a space--time wavefield $y: \mathbb{R}^2 \rightarrow \mathbb{R}$, we define its Fourier transform as 
\begin{equation}
    \hat{y}(\xi) = \int_{\mathbb{R}^2} y(\varsigma) e^{-2\pi i \xi^\transp \varsigma} d\varsigma,
\end{equation}
where $\xi$ is a wavenumber--frequency vector and decomposes into a wavenumber $k$ and frequency $\omega$ through $\xi = (k, \omega)^\transp$.

\begin{figure}
    \centering
    \includegraphics[width=0.75\linewidth]{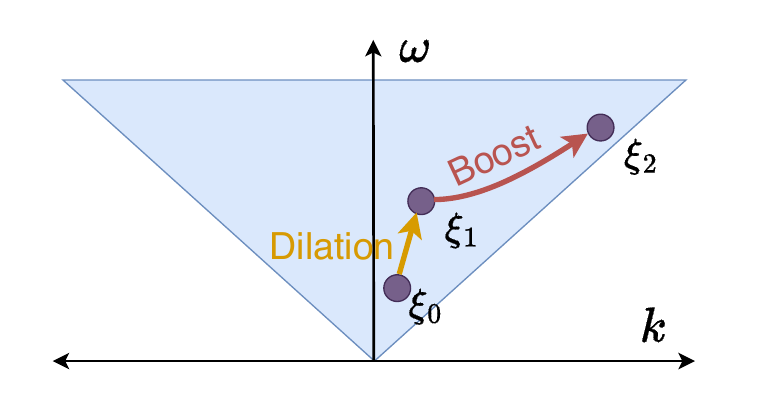}
    \caption{The action of the Weyl group on a 2D Fourier coordinate $\xi_0$ in the positive far-field cone. A dilation operator $D_a$ scales the coordinate isotropically to $\xi_1$, and a boost operator $B_\theta$ scales the coordinate anisotropically via hyperbolic rotations (boosts), resulting in the coordinate $\xi_2$. Note that the cone boundary can only be reached with an infinite boost $\theta \to \infty$.}
    \label{fig:cone}
\end{figure}

The action of $M_{a,\theta}$ in the wavenumber--frequency domain is illustrated in Figure~\ref{fig:cone}.
Note that these mappings naturally divide the wavenumber--frequency plane into two (double) cones: one for which $|\omega| < |k|$, called the \emph{near-field cone}, and one for which $|\omega| > |k|$, called the \emph{far-field cone}. In acoustics, the near- and far-field cones contain evanescent and propagating waves, i.e., phase speeds smaller or greater than the sound speed~\cite{Williams1999}. Indeed, by applying $M_{a,\theta}$ to a point in one of these cones, we can reach all other points in that cone.

We are now prepared to define the continuous boostlet function. 
\begin{definition}[Continuous boostlet]
Given a space--time translation vector $\tau \in \mathbb{R}^2$, a dilation factor $a \in \mathbb{R}^+$, and a boost parameter $\theta \in \mathbb{R}$, we define the boostlet function $\psi_{a,\theta,\tau}(\varsigma) \in L^2(\mathbb{R}^2)$ as~\cite{zeaBoostlets}
\begin{equation}
    \psi_{a,\theta,\tau}(\varsigma) = a^{-1} \psi\left(M_{a,\theta}^{-1} (\varsigma - \tau)\right),
\end{equation}
for some mother boostlet $\psi \in L^2(\mathbb{R}^2)$.
\end{definition}
The Fourier transform of the boostlet function is then given by~\cite{zeaBoostlets}
\begin{equation}
    \hat{\psi}_{a,\theta,\tau}(\xi) = e^{-2\pi i \tau^\transp \xi} \hat{\psi}(M_{a,\theta}^\transp \xi).
\end{equation}
The mother boostlet $\psi$ is chosen as a tensor product of Meyer-like wavelets parametrizing the near-field cone into dilations and boosts~\cite{zeaBoostlets}.

Given that dilation--boost matrices preserve the near- and far-field cone, a near-field mother boostlet $\psi$ will result in a family of boostlets all in the near field.
We will therefore adopt the convention that $\hat{\psi}(\xi)$ is supported in the near-field cone, i.e. $\hat{\psi}(\xi) = 0$ for $\xi = (k, \omega)$ such that $|k| < |\omega|$.
This near-field mother boostlet can then be transformed into a far-field mother boostlet by defining
\begin{equation}
    \psi^*(x, t) = \psi(t, x).
\end{equation}
This function will then have the property that $\hat{\psi}^*(\xi) = 0$ for all $|k| > |\omega|$, and is thus supported in the far field.
We can now define a family of far-field boostlets as in the near-field case by taking
\begin{equation}
    \psi^*_{a,\theta,\tau}(\varsigma) = a^{-1} \psi^*\left(M_{a,\theta}^{-1} (\varsigma - \tau)\right).
\end{equation}
The group defined in Definition 1 maps interior points of the cone to its boundary only in the asymptotic regime $\theta\rightarrow\infty$, thus necessitating the introduction of a scaling function to cover the remaining points in $\mathbb R^2$~\cite{zeaBoostlets} (see Sec.~\ref{sec:IIB}). 

\subsection{Decomposition and reconstruction}
\label{sec:IIB}

To decompose a wavefield $y \in L^2(\mathbb{R}^2)$, we simply compute its inner products with the near-and far-field boostlets
\begin{equation}
    B_{a,\theta,\tau} y = \langle y, \psi_{a,\theta,\tau} \rangle \quad \text{and} \quad B_{a,\theta,\tau}^*y = \langle y, \psi^*_{a,\theta,\tau} \rangle
\end{equation}
for all $a \in \mathbb{R}^+$, $\theta \in \mathbb{R}$, and $\tau \in \mathbb{R}^2$.
We call these the \emph{boostlet coefficients} of $y$.
If the Fourier transform $\hat\psi$ of the mother boostlet is band-limited in the near-field cone, it can be shown that $y$ may be recovered from its boostlet coefficients through~\cite{zeaBoostlets}
\begin{equation}
    y(\varsigma) = \int\limits_{\mathclap{\mathbb{R}^+ \times \mathbb{R} \times \mathbb{R}^2}} B_{a,\theta,\tau} \, y \, \psi_{a,\theta,\tau}(\varsigma) + B_{a,\theta,\tau}^* \, y \, \psi^*_{a,\theta,\tau}(\varsigma) \frac{da d\theta d\tau}{a^3}.
\end{equation}

Note that in the above, we have decomposed the original signal $y$ over all scales $a \in \mathbb{R}^+$.
For practical purposes, it is often useful to limit the maximum scale to some value $A > 0$ and, therefore, limit the decomposition to scales in the interval $(0, A]$.
The remaining information can then be recovered by a scaling function $\phi$ which satisfies by
\begin{equation}
    |\hat\phi(\xi)|^2 = 1 - \int\limits_{\mathclap{[0, A] \times \mathbb{R}}} |\hat\psi_{a,\theta,0}(\xi)|^2 + |\hat\psi^*_{a,\theta,0}(\xi)|^2 \frac{da d\theta}{a},
    \label{eq:scalingfun}
\end{equation}
which is translated to yield $\phi_\tau(\varsigma) = \phi(\varsigma-\tau)$ and decomposes $y$ into the \emph{scaling coefficients}
\begin{equation}
    Sy(\tau) = \langle y, \phi_\tau \rangle
\end{equation}
for $\tau \in \mathbb{R}^2$.
The original signal can then be recovered by the formula
\begin{equation}
    \begin{split}
        &y(\varsigma) = \int\limits_{\mathclap{\mathbb{R}^2}} Sy(\tau) \phi_\tau(\varsigma) d\tau \\
        &\quad + \int\limits_{\mathclap{[0,A] \times \mathbb{R} \times \mathbb{R}^2}} B_{a,\theta,\tau}\,y \, \psi_{a,\theta,\tau}(\varsigma) + B^*_{a,\theta,\tau} \, y \, \psi^*_{a,\theta,\tau}(\varsigma) \frac{da d\theta d\tau}{a^3}
    \end{split}
    \label{eq:reconstruct}
\end{equation}

In practice, the boostlet and scaling coefficients are sampled on a finite set of parameters $a$, $\theta$, and $\tau$.
The original function is then approximated by sums that approximate the above integrals. Examples of boostlet functions can be found in~\cite{zeaBoostlets}.


\section{Statistics of $n$-term approximations}
\label{sec:III}

To study the sparsity properties of the boostlet transform, we consider the $n$-term approximations generated by this transform, as well as two other representations: Daubechies45 wavelets~\cite{Daubechies1992} and cone-adapted shearlets~\cite{hauserShearlets}.
In other words, for each representation, we sort the coefficients $a_k$ by decreasing magnitude, truncate the sequence to $n$ coefficients, resulting in the sequence $a_{1:n}$, and use these to reconstruct a signal (e.g., using Eq.~\eqref{eq:reconstruct} in the case of the boostlet representation).
All three methods use $N_a = 2$ decomposition scales, and the boostlets have $N_\theta = 7$ decomposition boosts per scale. 

The analysis is performed on three wavefields measured in three different rooms at KTH with an omnidirectional microphone moved $100$ times and an omnidirectional sound source.
The sound pressure field is sampled every $3$~cm with a sampling frequency of $11.25$~kHz.
More details on the measurement experiment can be found in~\cite{Zea2019}.
Of these three room datasets, we pick $1000$ space--time windows of dimensions $100 \times 100$ by choosing a random starting time at all microphone positions (i.e., shifting the sound field across various time windows). 

\begin{figure}
    \centering
    \includegraphics[width=0.85\linewidth]{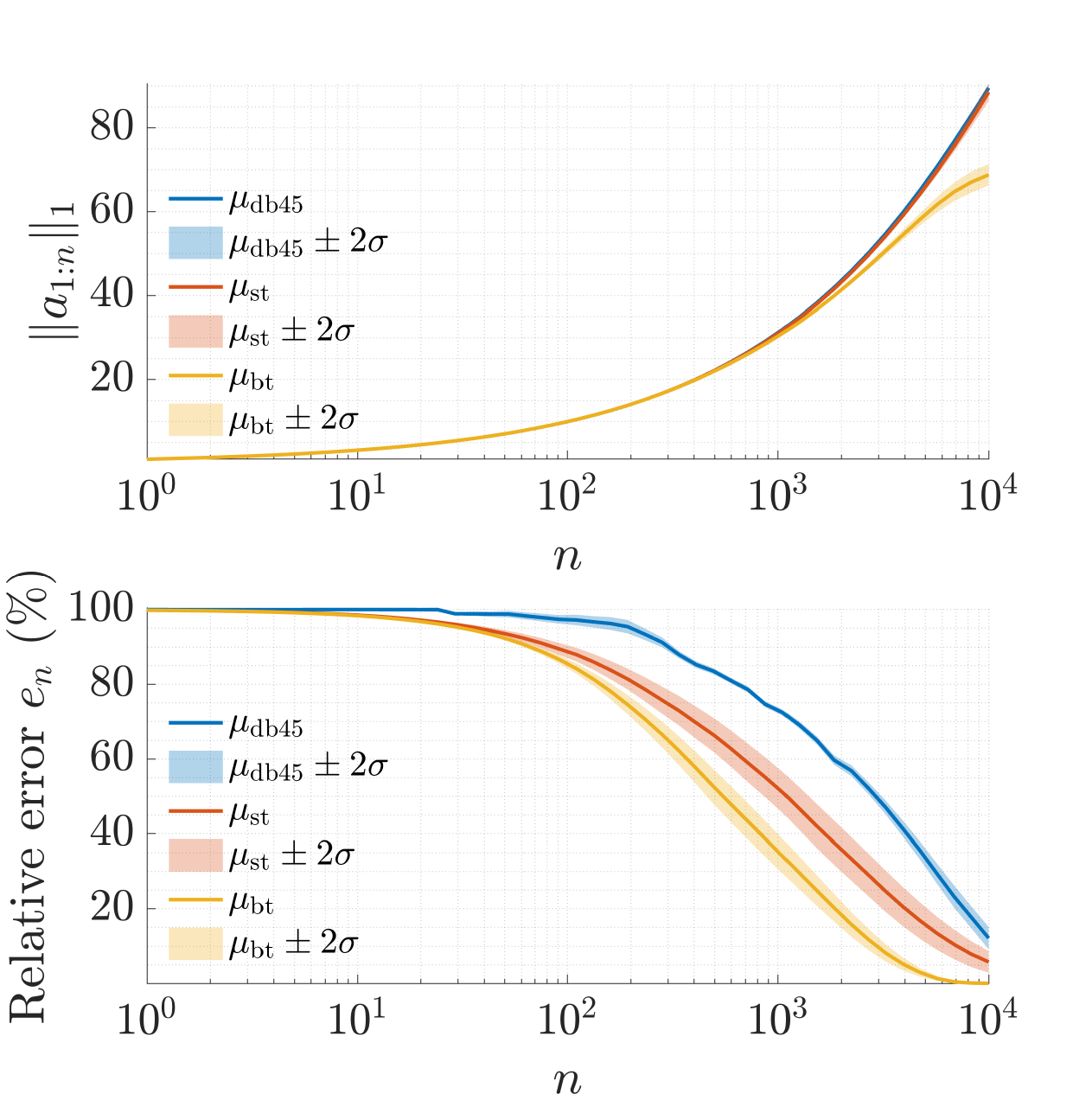}
    \caption{Statistics of $\ell_1$-norms and approximation errors, mean $\mu_\bullet$ with $95\%$ confidence intervals, of $n$-term reconstructions using Daubechies45 wavelets ($\bullet = \mathrm{db}45$), shearlets ($\bullet = \mathrm{st}$), and boostlets ($\bullet = \mathrm{bt}$) for $1000$ wavefields. (Top) $\ell_1$-norms against number of truncated coefficients $n$. (Bottom) Relative $n$-term approximation error $e_n$ against number of truncated coefficients $n$.}
    \label{fig:stats}
\end{figure}

For each wavefield $y$, we let $y_n$ denote its $n$-term approximation in some representation. 
Let us first consider the $\ell_1$-norm $\| a_{1:n} \|_1$ of these reconstructions, seen in Figure~\ref{fig:stats}(top).
The $\ell_1$-norms of the $n$-term boostlet approximations are lower than those of the corresponding wavelet and shearlet approximations.

Another perspective is offered by considering the reconstruction error of the $n$-term approximations.
Figure~\ref{fig:stats}(bottom) shows the  the relative $\ell_2$ error $e_n = \| y - y_n \|_2^2/\| y \|_2^2 \times 100\%$ for the different representations.
Here, we see that the boostlet representation achieves higher accuracy with significantly fewer coefficients than shearlets and wavelets.
It is important to note here, however, that the cone-adapted shearlets perform better than the wavelets, indicating that they provide a more natural representation for these types of signals.

\section{Denoising experiments via hard thresholding}

Given a clean wavefield signal $y$, we consider a noisy observation $y_\varepsilon = y + \varepsilon$, where $\varepsilon$ is some realization of a zero-mean noise distribution.
The goal is now to recover $y$ given $y_\varepsilon$.
One way to achieve this is to decompose $y$ in some representation, obtaining coefficients $a_k$ and then thresholding those coefficients, i.e., setting
\begin{equation}
    b_k = \begin{cases} a_k, & \text{if } |a_k| \ge \gamma, \\ 0, & \text{if } |a_k| < \gamma, \end{cases}
\end{equation}
for some threshold $\gamma > 0$~\cite{Mallat2008}.
The idea here is that if the representation is sparse for a certain class of signals (such as acoustic wavefields), it will concentrate most of its energy in a small number of coefficients, which will thus be large in magnitude.
However, the representation of the noise realization $\varepsilon$ is expected to be spread over many coefficients, which in turn will be of smaller magnitude.
As a result, the thresholding process will mostly zero out coefficients due to noise while preserving those in the desired class of signals.

\begin{figure*}[!ht]
    \centering
    \includegraphics[trim={4cm 1.5cm 4cm 2cm},width=0.8\textwidth]{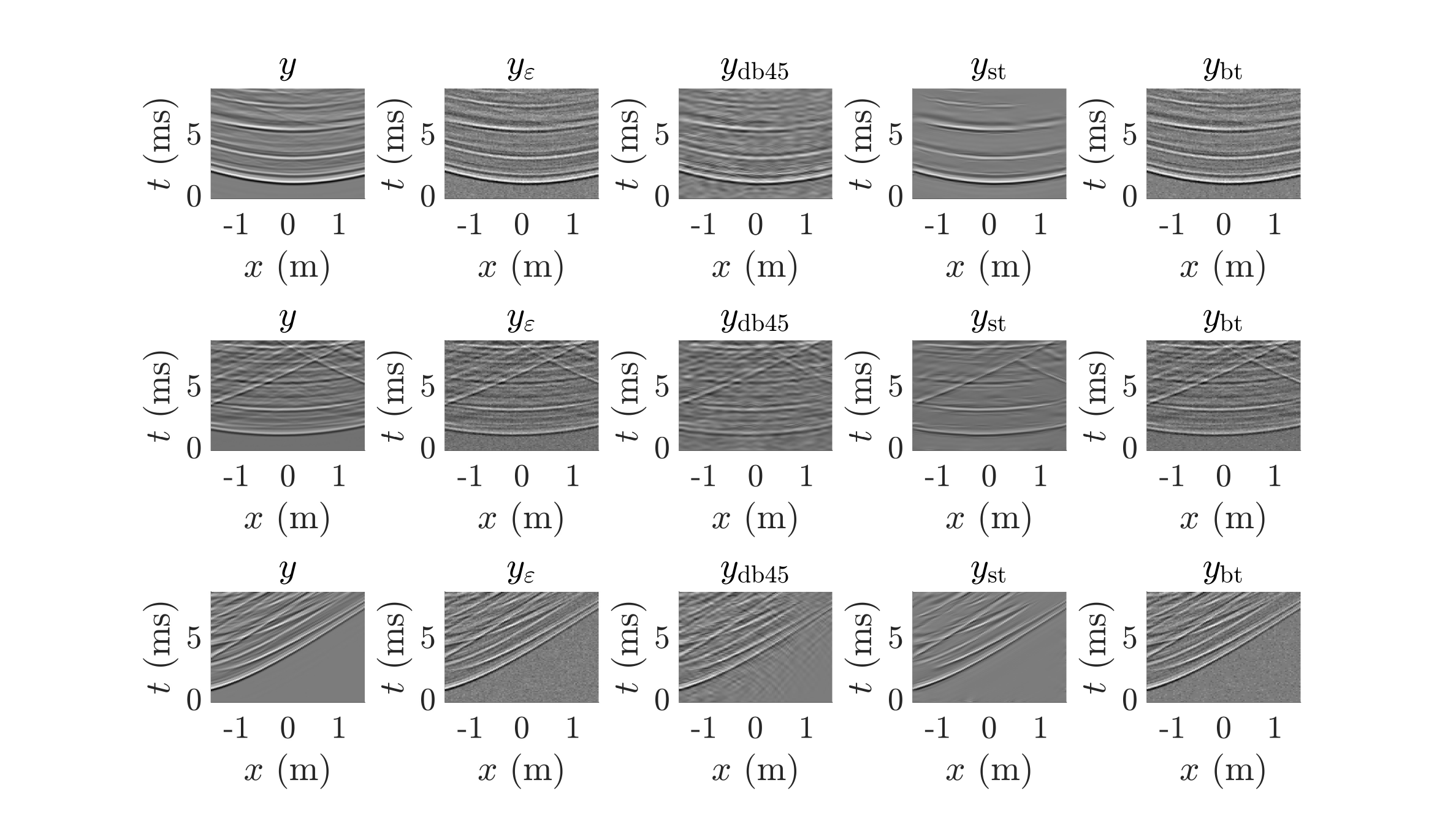}
    \caption{Denoising three spatiotemporal wavefields (rows) with $\mathrm{SNR} = 10$~dB. Left-most column: Ground-truth (noise-free) wavefield, $y$. Center-left column: Noisy wavefield, $y_\varepsilon = y + \varepsilon$. Center column: Denoised wavefield with Daubechies 45 wavelets, $y_\mathrm{db45}$. Center-right column: Denoised wavefield with cone-adapted shearlets, $y_\mathrm{st}$. Right-most column: Denoised wavefield with boostlets, $y_\mathrm{bt}$.}
    \label{fig:denoise1}
\end{figure*}

\begin{figure}
    \centering
    \includegraphics[width=\linewidth]{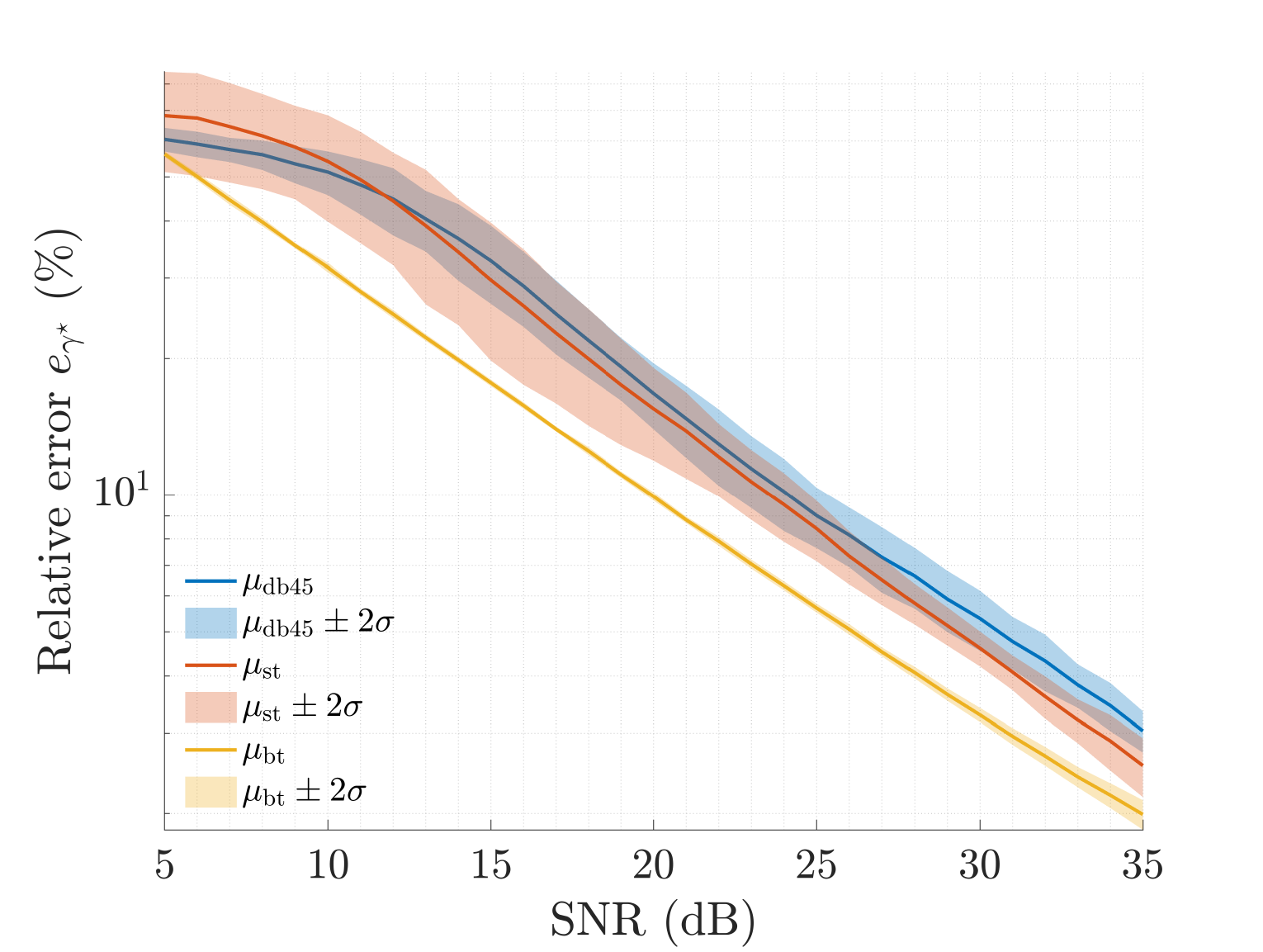}
    \caption{Statistics of reconstruction errors, $e_{\gamma^\star}$, with mean $\mu_\bullet$ and $95\%$ confidence intervals, against signal-to-noise ratio (SNR) for hard-thresholded reconstructions using Daubechies45 wavelets ($\bullet = \mathrm{db}45$), shearlets ($\bullet = \mathrm{st}$), and boostlets ($\bullet = \mathrm{bt}$) for $100$ wavefields.}
    \label{fig:denoise2}
\end{figure}

In this study, we investigate the influence of the signal-to-noise ratio (SNR) in dB, defined as $\mathrm{SNR} = 10 \log_{10} ( S / N )$, where $S$ and $N$ are the signal power and noise power, respectively. The SNR is varied between $5$ and $35$~dB, and $100$ randomly selected wavefields (see Sec.~\ref{sec:III}) are thresholded for each SNR value and for each representation: Daubechies45 wavelets, shearlets, and boostlets. A set of $100$ thresholds $\gamma$ is chosen for wavelets $[\sqrt{N},10\sqrt{N}]$, shearlets $[0.1\sqrt{N},5\sqrt{N}]$, and boostlets $[0.5\sqrt{N},\sqrt{N}]$. The optimal threshold, $\gamma^\star$, is obtained via the L-curve~\cite{Calvetti2004} parameter selection method. The L-curve is plotted as $\rho(\gamma) = \log \| y(\gamma) - y \|_2$ vs. $\eta(\gamma) = \log \| b(\gamma) \|_1$, and its point of maximum curvature corresponds to $\gamma = \gamma^\star$. Finally, the relative error for the thresholded reconstruction is computed as $e_{\gamma^\star} = \| y - y(\gamma^\star) \|_2^2/\| y \|_2^2 \times 100\%$. 

The reconstruction of three wavefields with wavelets, shearlets, and boostlets is shown in Fig.~\ref{fig:denoise1} for $\mathrm{SNR} = 10$~dB. The reconstruction errors for each wavefield (row) are summarized as follows. Daubechies45 wavelets yield relative reconstruction errors of $48.12\%$, $52.45\%$, and $50.34\%$, shearlets yield errors of $46.07\%$, $50.43\%$, and $41.67\%$, and boostlets yield errors of $31.09\%$, $31\%$, and $30.18\%$. 

Finally, the relative reconstruction errors $e_{\gamma^\star}$ against SNR are plotted in Fig.~\ref{fig:denoise2} for $100$ wavefields. It can be observed that the relative error of the boostlets, $\mu_\mathrm{bt} \pm 2\sigma$, is lower than that of wavelets, $\mu_\mathrm{db45} \pm 2\sigma$, and shearlets, $\mu_\mathrm{st} \pm 2\sigma$. Additionally, the spread of the confidence intervals is wider for wavelets and shearlets than for boostlets. This demonstrates that boostlets offer a representation more natural to denoise wavefield data. 

\section{Conclusion}
Our findings confirm that the boostlet decomposition offers a more compact representation of acoustic wavefields compared to conventional multi-scale transforms, such as wavelets and shearlets.
This manifests in higher-accuracy reconstruction from fewer coefficients and robust denoising performance through hard thresholding.
Altogether, these results underscore the natural role of boostlets for representing wavefields in unified space--time.

\section*{Acknowledgment}

The authors appreciate the discussions with Eric Brandão, Oriol Guasch, and Marc Arnela. 


\bibliography{example.bib}{}
\bibliographystyle{IEEEtran}

\end{document}